# Decentralized creation of academic documents using a Network Attached Storage (NAS) server

Johannes Wilm[1], Afshin Sadeghi[2], Christoph Lange[3], Philipp Mayr[4].

**Abstract.** Scholarly document creation continues to face various obstacles. Scholarly text production requires more complex word processors than other forms of texts because of the complex structures of citations, formulas and figures. The need for peer review, often single-blind or double-blind, creates needs for document management that other texts do not require. Additionally, the need for collaborative editing, security and strict document access rules means that many existing word processors are imperfect solutions for academics. Nevertheless, most papers continue to be written using Microsoft Word (Sadeghi et al. 2017). We here analyze some of the problems with existing academic solutions and then present an argument why we believe that running an open source academic writing solution for academic purposes, such as Fidus Writer, on a Network Attached Storage (NAS) server could be a viable alternative.

1. Dr. Johannes Wilm, GESIS–Leibniz Institute for the Social Sciences, Unter Sachsenhausen 6-8, 50667 Köln, E-Mail: mail@johanneswilm.org
2. Afshin Sadeghi, University of Bonn, Römerstraße 164, 53117 Bonn, E-Mail: sadeghi@cs.uni-bonn.de
3. Dr. Christoph Lange, University of Bonn, Römerstraße 164, 53117 Bonn und Fraunhofer IAIS, Schloss Birlinghoven, 53754 Sankt Augustin, E-Mail: langec@cs.uni-bonn.de
4. Dr. Philipp Mayr, GESIS–Leibniz Institute for the Social Sciences, Unter Sachsenhausen 6-8, 50667 Köln, E-Mail: philipp.mayr@gesis.org

## Issues with existing approaches

### Issues with Microsoft Word and other general word processors for academics

Microsoft Word has the advantage that users are familiar with its easy to use What-You-See-Is-What-You-Get (WYSIWYG) interface. However, articles authored in Word have the problem that they lack semantic information, which means that conversion for final publication into other formats will be more difficult and will require human intervention. Because the conversion process is imperfect, there is also the chance of loss of information or misinterpretation on the part of the human executing the conversation. The same is true for open source alternatives with a similar user interface and workflow, such as LibreOffice or OpenOffice.org.

Additional problems occur if one needs to collaborate on a text among several writers: Collaborative editors such as Google Docs or Microsoft Office 365 Online place the documents on servers outside of the control of the user, and potentially confidential information is shared with companies operating servers in countries where national spy agencies are working hard on obtaining copies of all digital information that enters and leaves the country. Their work is even easier if they can find all information on servers operated by just a small number of companies.

### Issues with web services targeting academic writers

Some new online writing services have appeared in recent years[5] to target academic writers specifically. These editors handle citations, formulas and the like, and conversions to final output formats will therefore require less or no human intervention.

While these editors may have solved much of the conversion issues, in so far as they are closed source applications hosted by a single company, they have the same security issues as Google Docs and Microsoft Office 365 Online. Open source alternatives, such as Fidus Writer and ShareLaTeX, in theory have less of this issue, as the application can be installed on a secure server. In practice, most users will not have their own regular server, so that this option is not really accessible to them. Should they hire server space from a third party, they may have more control over which country their data will be stored in, but they will still be exposing their data to the company operating the server.

---

[5] Among the academic text editing apps that have appeared are: Authorea, Fidus Writer, ShareLaTeX and WriteLaTeX/Overleaf

**Issues with decentralized document editing in a collaborative way**

An alternative for decentralized collaboration is one where the editing application is installed on the end user's computer. This solves the problem of the server, but it creates a number of other challenges:
- General installation processes are too complex for novice users. The fact that the developers of the editing software cannot know which OS their end users will be running makes it hard to give standardized installation instructions.
- Users today are often working on sections of the Internet behind routers with changing IP numbers on the internet. Permanent IP numbers are given out by Internet Service Providers (ISPs) only for an extra fee or not provided at all. Without any other aids, it is therefore somewhat tricky to connect two computers with each other if they are on different local networks. While there are ways around this problem accessible to IT professionals, it will likely be too complex for the average end user.
- If two users are collaborating on writing a document, but they are working at different times and cannot guarantee that either computer is on the Internet the entire time, merging changes becomes a problem. Even though merging mechanisms may find a way to automatically merge changes, there is no guarantee that the human language and argument described in the text still make sense if texts are not merged immediately and writers can be guaranteed that the version they are working on contains all the additions made by other collaborators.

Taken together, these points mean that a direct real-time collaboration setup cannot reasonably be established and run by computer novices without significant help from IT professionals.

## Decentralized document editing using a Network Attached Storage (NAS) Server

Another option is the installation of a small and local server on the side of one of the document editors. Even search giant Google, known for its various cloud services, recognizes the need for data stored locally on smaller servers as it offers its Google Search Appliance product for companies with large amounts of data[6]. NAS servers can fulfill a similar role, but scaled to the comparatively smaller amount of data needed within an academic text editing setting.

NAS servers solve some of the issues encountered when trying to run collaboration from the end user's computer: If one targets one specific NAS platform, the installation process can be simplified to a few clicks and filling in forms that even novice computer users can handle. The lack of a permanent IP address is made up by a dynamic DNS service offered by the vendor of the NAS server. As the installation and management procedures of NAS servers generally take place through web interfaces, they are made to be capable of serving at least basic web pages. As long as the NAS Server can be turned on and connected to the Internet constantly, it does not matter that the end users connect to the document at different times. Another alternative may be a mini-computer such as "Linux boxes" that are sold within a similar price range of around 90-200 Euros. Different from NAS servers, they are targeting more advanced users and do not always come with the same dynamic DNS service built-in.

The usage of NAS servers for this purpose creates some other challenges however, as the main purpose of the devices is that of a storage device for backup of files and not a general web server for real-time collaboration. CPU power and RAM are therefore somewhat limited. Also, the NAS represents an extra cost to the end user, which means that the more costly devices will not be an option in many cases.

## Test setup

In order to find out whether a NAS Server would be a practical alternative, we took a NAS device from the lower end of the spectrum – a Synology DS215J that had been running online for 2.5 years for backup purposes – and we tried to install Fidus Writer on it. Fidus Writer was chosen because it targets specifically computer novices in the humanities and social sciences who require a WYSIWYG user interface that still provides all the features needed for scientific text editing (Wilm and Frebel 2015). The DS215J has a 800 MhZ Marvell Armada 375 Dual Core CPU and 512 MB of RAM. Currently the successor version of the DS215J, the DS216J, sells for around 163 Euros (May 2017)[7]. The NAS was installed on a home network connected to the internet with a 51.4 MBit (Down)/10 MBit (Up) DEutsche Telekom connection, in northern Germany. Tests were performed from southern Sweden with a 55.5 MBit (Down)/47 MBit (up) Telecom 3 Sverige AB connection. Total air distance between NAS and connected computer was 271 km and this part of the world is generally known for having a good connection quality.

---

6     https://enterprise.google.com/search/products/gsa.html

The installation process was relatively easy. However, we decided against trying to package the app properly for this initial test, as the purpose of the test was to see whether the hardware limits of the NAS would be an issue for speed or connectivity. Installing the application directly via the command line onto the Linux version already running on the device also proved challenging, as header files for libraries, etc. were missing. In the end we decided that the fastest way of arriving at our goal was to install a Debian *chroot* environment for which there was a community-built package available. The installation instructions provided with Fidus Writer[8] are written for Ubuntu 16.04, and these proved to be close to, but not the exact same as what was needed for Debian Jessie. Most notably did we need to install a newer version of Node.js than what the packaging system provided us with. We were then able to set the system up, connecting it even with a MariaDB database provided by another standard package on the NAS. The NAS was also able to reprogram the router to give us access to the port we decided to run our application on from the outside. The entire installation was done remotely without physical access to the NAS.

## Test results

Our tests showed that at five clients could be connected to the NAS servers simultaneously and read/write the same document without any noticeable anomalies. We did not attempt to connect with more than five clients, as this number seemed more than sufficient for our purposes. While there were a few situations where the page would not load entirely the first time and it had to be reloaded, we attribute this to internet connection issues and not the NAS. The running of the combined Tornado/Django server for that makes out the backend of Fidus Writer and which is needed especially for the collaboration part, did not present a challenge to the NAS server.

---

7 https://www.amazon.de/gp/offer-listing/B01BVDJGPE/ref=sr_1_1_olp?ie=UTF8&qid=1495437805&sr=8-1&condition=new

8 https://github.com/fiduswriter/fiduswriter/blob/3.1.0/README.md

Other parts of the editor – such as document export or import of citation sources in the BibTeX format – ran smoothly as well, but this was less of a surprise for us as we knew Fidus Writer had been programmed in client-heavy way, shifting most of the computing processes onto the client (browser) and only doing what is the minimum required on the server. Processes such as handling incoming document updates from clients and exporting/importing files are therefore implemented as much as possible in the browser and do not require server resources.

## Conclusion

Academic document production continues to be challenging, especially when dealing with confidential material and when wanting to write in a way that preserves semantic information to avoid problems in the later stages of the publication process. Running open source semantic editing software is challenging because not everyone has access to their own server or can trust companies running such servers for them. An installation of a semantic editing software on a NAS server seems in many cases to be a viable alternative, as our tests running Fidus Writer on a Synology DS215J showed. Client-heavy applications such as Fidus Writer are well-suited for the job, as they will only require the minimal amount necessary from the NAS-servers and perform all other calculations in the browser. Packaging Fidus Writer as an app for the Synology system remains to be done before usability studies of the setup can commence.

## Acknowledgement

This work was funded by DFG, grant no. SU 647/19-1 and AU 340/9-1; the OSCOSS project at GESIS and University Bonn (Mayr and Lange 2017). We would also like to thank other project participants and Fidus Writer participants who made the current version of Fidus Writer possible. Among these are Mana Azamat, Niloofar Azizi, Daniel Frebel, Babak Hashemi, Firas Kassawat, Takuto Kojima, Aleksandr Korovin, Fakhri Momeni, and Anne Wittorf Kojima.

# Reviews

## Review 1

Author: Sarven Capadisli, University of Bonn

The article makes sound points on why a NAS server would be useful for some types of users. Would something like this interest individuals to have their own (at home) or would it be simpler for teams and labs having one where they share?

Have you already or plan to do a survey on which domain of scholars and how many would use a NAS server for writing academic articles? Or would this more of an challenge to deploy an FW app for NAS systems?

>potentially confidential information is shared…

I don't quite understand the problem or who that may particularly effect.

How much total disk space is used on the NAS including Fidus Writer installation?

## Answer 1

Author: Johannes Wilm, GESIS–Leibniz Institute for the Social Sciences

> Would something like this interest individuals to have their own (at home) or would it be simpler for teams and labs having one where they share?

My guess is that this would depend on how the individual journal operates. Some are heavily tied to institutional infrastructure and require for a NAS to be setup at a shared office. Others are more open to letting individuals use their own infrastructure and having a journal eeditor run such a NAS server from home. Running it from home will likely be the easiest technically speaking.

> Have you already or plan to do a survey on which domain of scholars and how many would use a NAS server for writing academic articles? Or would this more of an challenge to deploy an FW app for NAS systems?

This is our first investigation into the viability of a NAS-based solution. We are currently discussing into which direction to take it further. A survey on what kinds of editors would be interested in this is among the items we are currently discussing. The packaging of Fidus Writer as an app seems to be time-consuming and less of a technical challenge.

> >potentially confidential information is shared…

> I don't quite understand the problem or who that may particularly effect.

It is as if you give first-hand preview access to the companies running the server. The companies themselves may stick to a policy of not reading their clients content, but they could be ordered to share the content with an agency through a secret court order, etc. . If, on the other hand, you have the NAS running and everyone connected to it is connecting from the same country or the same region, the data would not be as easy to get for the same agencies and the big companies will have even fewer ways to get at this information.

> How much total disk space is used on the NAS including Fidus Writer installation?

This is difficult to say, as our current setup depends on the installation of an entire Debian distribution, whereas a packaged version would use the Linux version that comes bundled with the NAS. Also, there is the question of whether one should one count the size of the database package and other dependencies, which may have been installed already due to other systems that are being run on the NAS. Fidus Writer by itself takes around 235 MB.

**Review 2**

Author: Amy Guy, University of Edinburgh / MIT

This article addresses quite a specific problem: collaborative document editing without third-party servers. The reasons for not wanting to rely on Web servers, even if they're under the control of the document author, is fairly well explained and motivated.

Addressing authoring solutions to MS Word users is diving in the deep end, but a worthy goal as this reaches a lot of researchers. However, I question that the idea of installing software is too complicated for most people.. do you not think it's possible to get the installation process for FW to be as simple as for MSWord?

I realise that this article is not about FidusWriter, but rather collaborative document editing via a NAS, but it would be helpful to explain more about what FidusWriter does and how it is normally run, ie. not on a NAS, and some comparison.

The authors seem to feel strongly about data ownership, mentioning privacy potentially being violated by corporations and governments. It's not clear however that FidusWriter provides a publishing solution, only an authoring solution. That is, once an article is finished it seems that a PDF is exported and turned over to third parties anyway? Relatedly, I would love to know more about how the authors envision this fitting in with the rest of the scholarly communication process, as mentioned in the abstract, particularly peer-review. Could a similar setup be purposed to permit reviewers to control their review contents as well? I assume FidusWriter takes care of access control, citations, formulas and figures, though the article does not state that.

The article mentions that there are other web services for academic authoring, but doesn't name any. I'd be interested to know which ones the authors analysed. Is it only the ones in the footnote? In which case only Authorea and Overleaf are problematic in the ways mentioned for running on third-party servers. I don't really see evidence that the authors have performed a comprehensive search for alternatives in this space. There are certainly clientside document authoring applications which run on personal data stores, such as Laverna which can talk to RemoteStorage servers, and dokieli (which I'm pretty sure the authors have heard of) which can talk to LDP servers, not to mention 'decentralised Google Docs' type things like NextCloud and CosyCloud. It would be worth finding out how easily any of these could be extended to add collaborative editing if they don't have it already, since they have the decentralisation part covered.

Thus, perhaps it's worth comparing running FidusWriter on a NAS to running various generic personal datastores. It seems to me that FidusWriter rather ties the editing application with the storage, which prevents the user from easily switching applications. It's not clear if FW follows a standard protocol for data exchange with the server either. These aren't really the problems of the authors in this context, but it is a downside of a non-standard system even if it's open source and installed on the user's machine. Useful future work might be taking the idea of a NAS forward with other storage/server and client/application options.

That said, I would love to see a docker image prepared to easily set up FidusWriter on any server, if that doesn't already exist!

Limitations are appropriately mentioned as cost and power of NAS devices, though no specifics are given.

Finally, the authors assume a status-quo vision of academic authoring (eg. the "need" for blind peer review) and proceed from there with a decentralisation angle. More interesting would be a paragraph or two about how an approach like this is setting the stage for future advancement in the space. I like the idea of meeting in the middle; addressing immediate problems authors have with their current tooling, whilst laying some foundations for progressing in the direction opened up to us by Web technologies.

Minor comment: what is meant by "Linux boxes" (in quotes)? Are you getting at things like Raspberry Pis?

And I can't help but ask: did you collaboratively author this article using FidusWriter on a NAS? Some screenshots of doing so would add a whole new layer of credibility to your analysis

**Answer 2**

Author: Johannes Wilm, GESIS–Leibniz Institute for the Social Sciences

Thanks for the reviews, both of you!

> I question that the idea of installing software is too complicated for most people.. do you not think it's possible to get the installation process for FW to be as simple as for MSWord?

I don't think any of us want t speak in the name of all or the majority of users. For me, the assessment that installing software is too difficult for users is based on about 15 years of practical experience with social scientists and scientists in the humanities. Tryin to get social scientists to install LyX or similar tools, even with a rather painfree Windows installer, turned into an absolute nightmare for the involved, which they couldn't surpass until I personally took their computer and installed the software for them. Even then, their usage of the software only worked so long as they did not accidentally change some setting. Through communication with various people in the publishing field, this perception seems to be close to the experience of others as well. We do not have statistical data on whether the installation process itself is what kept users from using certain software, but the survey we cite does show that academic users do find certain features (citation management, etc.) useful, yet opt for solutions that do not provide this in a sensible way anyway.

Microsoft Word comes bundled with computers and we will likely never achieve something as simple, just because most users are not academics and will not want that. But the installation process on a NAS may be as simple as clicking a few places and writing a few letters into search boxes and alike.

> I realise that this article is not about FidusWriter, but rather collaborative document editing via a NAS, but it would be helpful to explain more about what FidusWriter does and how it is normally run, ie. not on a NAS, and some comparison.

We have written about Fidus Writer in earlier articles. Given that we are also the same people who have done a lot of work in Fidus Writer, I think we didn't write more about it this time because we felt there should be a limit to self-promotion. The point is taken though and in a next version we should probably include more of that.

> It's not clear however that FidusWriter provides a publishing solution, only an authoring solution. That is, once an article is finished it seems that a PDF is exported and turned over to third parties anyway? Relatedly, I would love to know more about how the authors envision this fitting in with the rest of the scholarly communication process, as mentioned in the abstract, particularly peer-review.

Fidus Writer can be connected with the Open Journal Systems (OJS) to form a complete publication writing and management solution. It's final output format can currently be ODT, DOCX, HTML, EPUB, PDF, LaTeX or the native Fidus file format. Creating new export filters is on the agenda, but we need a bit more information what other file formats are most useful.

> I assume FidusWriter takes care of access control, citations, formulas and figures, though the article does not state that.

Indeed, it does. This article only states this indirectly probably in order to avoid shameless self-promotion, but we should probably say this more explicitly in a future version.

> I don't really see evidence that the authors have performed a comprehensive search for alternatives in this space.

The document editing solutions mentioned are the leading solutions know today. We are aware that there are various experiments using clientside technology, such as dokieli. But as far as we could tell, their usage of clientside technology prevents them from providing real-time collaboration. We have mentioned this as a general issue of clientside technologies in the article, without mentioning specific editors.

From what we knew before, Dokueli is not capable of realtime collaboration. And as far as I can tell by looking at the mentioned systems and reading the latest draft spec of the remote storage spec, realtime collaboration will remain impossible due to architectural reasons, unless I am missing something. Laverna does not seem to work on realtime collaboration, and does not provide any features for academics as far as I can tell. The editors available in owncloud/nextcloud do not seem to be targeting scholarly communication. Integrating Fidus Writer with these cloud application could be a future project.

We are not trying to say that Dokieli and alike are not of interest to certain users. We are not claiming that everyone should use Fidus Writer or any other particular editors on a NAS server. Nor do we claim that everyone should be doing real-time collaboration. There are probably many reasons why certain users would want to connect a clientside editor with a storage solution somewhere else. All we are saying is that some users require realtime collaboration, citation management in a WYSYWIG-alike environment in an open source solution, and there seems to be no alternative right now. A comparison also with these other editors could be the theme of another paper.

> It seems to me that FidusWriter rather ties the editing application with the storage, which prevents the user from easily switching applications. It's not clear if FW follows a standard protocol for data exchange with the server either.

The ability to access the same data with different applications seems to be a major theme for the proponents of dokueli and the remote storage spec. For FIdus Writer the situation is complicated because the document is not only stored in a database. In addition, while opened, the Fidus Writer server needs to manage the flow of editing operations between clients connected to this document and these are only stored to the database at certain intervals or when all connections to a particular document have been closed.

Fidus Writer uses operations more similar to some other realtime collaboration apps by having replaced its own realtime collaboration system with ProseMirror as of version 3.0, and ProseMirror is being used by other appliations as well. This does however not mean that the various editor apps can interchange each other's backends, as there are many other complicating factors around this, and most other text editors are not focused on academic purposes, so they won't support citations, figures, formulas, etc. .

Being able to realtime collaborate on texts between Fidus Writer and another text editing app is not a major priority for us and will probably lie many years into the future. Adding more export and import filters to write and read to standardized file formats seems to be a more important improvement at this stage.

The communication Fidus Writer has with it's database does of course follow SQL conventions.

> That said, I would love to see a docker image prepared to easily set up FidusWriter on any server, if that doesn't already exist!

A user recently informed us that he had created a docker image for Fidus Writer: https://hub.docker.com/r/moritzf/fiduswriter/

> Finally, the authors assume a status-quo vision of academic authoring (eg. the "need" for blind peer review) and proceed from there with a decentralisation angle. More interesting would be a paragraph or two about how an approach like this is setting the stage for future advancement in the space. I like the idea of meeting in the middle; addressing immediate problems authors have with their current tooling, whilst laying some foundations for progressing in the direction opened up to us by Web technologies.

We have been discussing adding other types of reviews/feedback. We think that what this group is doing is very interesting, but due to time constraints on our part, we have focused for now on providing what is needed for the traditional workflow.

> Minor comment: what is meant by "Linux boxes" (in quotes)? Are you getting at things like Raspberry Pis?

No, not really. We just meant small and relatively inexpensive ccomputers that run linux such as this one: https://www.amazon.de/Celeron-Fanless-Support-included-Aluminum/dp/B01GBHBJYQ/ref=sr_1_3?ie=UTF8&qid=1495944070&sr=8-3&keywords=linux%2Bbox

These could be an alternative buyign choice for a consumer considering buying a NAS as they cost about the same. They have much strong processors and will be able to run Fidus Writer without a problem, but they do not have the builtin dynamic DNS, easy install system, etc. .

> And I can't help but ask: did you collaboratively author this article using FidusWriter on a NAS? Some screenshots of doing so would add a whole new layer of credibility to your analysis

No, we authored this using Microsoft Word and sent the file back and forth.

Just kidding. We did write this collaboratively using FIdus Writer, but we used the instance Gesis has setup in its server park. This was mainly out of convenience as we had used it for another paper just prior to that and knew that the server was properly configured. We had previously talked about looking at setups on NAS servers and thought we should try it out when we heard about this call a few weeks ago. It was all fairly last minute for us and we were rather surprised that we were able to get it running at all.